\documentclass{ar2e}
\usepackage[dvips]{graphicx}
\newcommand{\simgt}{\lower.5ex\hbox{$\; \buildrel > \over \sim \;$}}
\newcommand{\simlt}{\lower.5ex\hbox{$\; \buildrel < \over \sim \;$}}
\newcommand{\bm}[1]{\mbox{{\it \boldmath$#1$}}}

\begin{document}

\baselineskip=15pt

\input epsf.sty

\jname{Ann. Rev. Nuclear and Particle Science}
\jyear{2008}
\jvol{58}
\ARinfo{1056-8700/97/0610-00}

\title{Weak Gravitational Lensing and its Cosmological Applications}

\markboth{Hoekstra \& Jain}{Cosmology and Gravitational Lensing}

\author{Henk Hoekstra 
\affiliation{Department of Physics and Astronomy, University of Victoria, 
3800 Finnerty Road, Victoria, BC, V8P 5C2, Canada. Email:hoekstra@uvic.ca}
Bhuvnesh Jain
\affiliation{Department of Physics and Astronomy, University of
  Pennsylvania, Philadelphia, PA 19104. Email:
  bjain@physics.upenn.edu}}


\begin{abstract}

Weak gravitational lensing is a unique probe of the dark side of the
universe: it provides a direct way to map the distribution of dark
matter around galaxies, clusters of galaxies and on cosmological
scales.  Furthermore, the measurement of lensing induced
distortions of the shapes of distant galaxies is a powerful probe of
dark energy. In this review we describe how lensing
measurements are made and interpreted. We discuss 
various systematic effects that can hamper progress and how they may
be overcome.  We review some of the recent results in weak lensing 
by galaxies, galaxy clusters and cosmic shear and discuss the
prospects for dark energy measurements from planned surveys.

\end{abstract}

\maketitle

\section{Introduction}

The deflection of light rays by intervening structures, a phenomenon
referred to as gravitational lensing, provides astronomers with a
unique tool to study the distribution of dark matter in the
universe. Unlike other observational probes, the lensing effect
provides a direct measure of the mass, irrespective of the dynamical
state of the lens.  A well publicized recent example is the study of
the merging `Bullet' cluster of galaxies, in which a lensing study showed
that the dark matter is displaced from the bulk of the baryonic
mass \cite{Clowe06}. Such measurements can provide important insights
into the properties of dark matter.

A number of applications of gravitational lensing have proven
important for observational cosmology. When the lens is sufficiently
strong, multiple images of the same source can be observed. If the source
is variable, time delays between the variation of the images can be
determined. An example of the applications of strong lensing is 
in the use of time delays to estimate the Hubble constant, 
provided a good model for the lens can be derived
(see e.g. \cite{Refsdal64,Koopmans03}). In this review, however, the focus
will be on applications of weak gravitational lensing 
that can help us understand the properties of
dark energy and dark matter on cosmological scales. 

Weak lensing refers to the shearing of distant galaxy images due to
the differential deflection of neighboring light rays. The signal is
small, typically inducing an ellipticity of order 1\%. While this is 
negligible compared to the intrinsic shape of individual galaxies, it
can be measured statistically using the coherence of the lensing shear
over the sky.  
In the past two decades it has become possible to measure these
subtle changes to study the distribution of dark matter in the
universe. The first measurements used lensing by galaxy clusters; 
more recently cosmic shear measurements have been made in ``blank
fields'', without using any knowledge of foreground structures. 
A number of topics related to galaxy and cluster lensing 
are discussed in \S5, but the main focus of this review is cosmic
shear, i.e. lensing by large-scale structure in the universe. 

The reason for the recent popularity of cosmic shear is the fact that
the signal is a direct measure of the projected matter power spectrum
over a redshift range determined by the lensed sources (see e.g.,
\cite{Blandford91, Kaiser92}).  This straightforward interpretation of
the signal is rather unique in the tools available for cosmology, and
it potentially enables the determination of cosmological parameters with
high precision.  Lensing measurements are not only sensitive to the
geometry (similar to distance measures such as type Ia supernovae or
baryonic acoustic oscillations), but also provide measures of the
growth of large-scale structure that test gravity on cosmological
scales. These features make cosmic shear one of the most powerful
probes of dark energy and modified gravity theories
\cite{Albrecht06,Peacock06}, albeit an observationally challenging
one.

This review focuses on the methods and principal challenges in 
the cosmological applications of weak lensing. We review lensing
theory but the focus is on measurements, current surveys and prospects
for planned surveys in the coming decade. We refer the interested
reader to reviews with a more detailed treatment of many
other aspects of lensing \cite{Bartelmann99,Mellier99,Refregier03b,Munshi06}. 

In \S2, we describe the key steps in the measurement and in \S3
we discuss the interpretation of cosmological weak lensing. 
In \S4 we review the primary systematic errors, as well as ways to 
deal with them. We highlight some of the current results in
cosmological weak lensing in \S5. In
\S6 we discuss lensing by galaxies and galaxy clusters. We conclude
in \S7 with a discussion of prospects for the coming decade.

\section{How to measure shear}

\subsection{Weak lensing basics}

Massive structures along the line of sight deflect photons originating from
distant galaxies. If the source is small, the effect is a (re)mapping
of $f^{\rm s}$, the source's surface brightness distribution (see
Reference \cite{Bartelmann99} for more details):

\begin{equation}
f^{\rm obs}(\theta_i)=f^{\rm s}({\cal A}_{ij}\theta_j),
\end{equation}

\noindent where ${\cal A}$ is the distortion matrix (the Jacobian
of the transformation)

\begin{equation}
{\cal A}=\frac{\partial(\delta\theta_i)}{\partial \theta_j}
=(\delta_{ij}-\Psi_{,ij})=
\left(
    \begin{array}{cc}
        1-\kappa-\gamma_1 & -\gamma_2 \\
        -\gamma_2        & 1-\kappa+\gamma_1 \\
    \end{array}
\right)
\label{distort}
\end{equation}

\noindent where we have introduced the two-dimensional lensing
potential $\Psi$, and where $\Psi_{,ij}\equiv\partial^2 \Psi/{\partial
\theta_i\partial\theta_i}$. The lensing convergence $\kappa$ is a
scalar quantity and is given by a weighted projection of the mass
density fluctuation field:
\begin{equation}
\kappa(\bm{\theta})=\frac{1}{2}\nabla^2\Psi(\bm{\theta})
=\int\!\!d\chi W(\chi) 
\delta[\chi, \chi\bm{\theta}],
\label{eqn:kappa}
\end{equation}
with the Laplacian operator $\nabla^2$ defined using the flat sky
approximation as $\nabla^2\equiv
\partial^2/{\partial\bm{\theta}^2}$ and $\chi$ is the comoving
distance (we have assumed a spatially flat universe).  Note that $\chi$
is related to redshift $z$ via the relation $d\chi=dz/H(z)$, where $H(z)$ is
the Hubble parameter at epoch $z$.  
The lensing efficiency function $W$ is given by
\begin{equation}
W(\chi)=\frac{3}{2}\Omega_{m0}H_0^2 a^{-1}(\chi)
\chi \int\!\!d\chi_s~ 
n_s(\chi_s) \frac{\chi_{\rm s}-\chi}{\chi_s},
\label{eqn:weightgl}
\end{equation}
where $n_s(\chi_s)$ is the redshift selection function of source galaxies
and $H_0$ is the Hubble constant today 
($H_0=100h{~\rm km~s}^{-1}{\rm Mpc}^{-1}$). 
If all source galaxies are at a single redshift
$z_s$, then $n_s(\chi)=\delta_D(\chi-\chi_s)$.  

In Equation~\ref{distort} we introduced the components of
the complex shear $\bm{\gamma}\equiv\gamma_1+i\gamma_2$, which can
also be written as $\bm{\gamma}= \gamma\ {\rm exp}(2i\alpha)$, where $\alpha$
is the orientation angle of the shear. The Cartesian components of the
shear field are related to the lensing potential through
\begin{equation}
\gamma_{1}=\frac{1}{2}(\Psi_{,11}-\Psi_{,22})\hspace{1em}{\rm and}\hspace{1em}
\gamma_2=\Psi_{,12},
\label{sheardef}
\end{equation}

In the weak lensing regime, the convergence gives the magnification
(increase in size) of an image and the shear gives the ellipticity
induced on an initially circular image. Under the assumption that
galaxies are randomly oriented in the absence of lensing, the strength
of the tidal gravitational field can be inferred from the measured
ellipticities of an ensemble of
sources (see \S4.2 for a discussion of intrinsic alignments). In the
absence of observational distortions, the observed ellipticity $e^{\rm
obs}$ is related to its unlensed value $e^{\rm int}$ through
\cite{Seitz97,Bartelmann99}:

\begin{equation}
e^{\rm obs}=\frac{e^{\rm int}+ \gamma}{1+\gamma^* e^{\rm int}},
\label{ave}
\end{equation}
where $e \simeq \left[(1-b/a)/(1+b/a)\right]{\rm exp}(2i\alpha)$ for
an ellipse with major and minor axes $a$ and $b$, respectively, and
orientation angle $\alpha$. $\gamma^*$ is the complex conjugate of the
lensing shear. The average value of $e^{\rm obs}\approx \gamma$ in the
weak lensing regime. To be more precise, the observable is the reduced
shear $\gamma/(1-\kappa)$. Hence, the unbiased measurement of the
shapes of background galaxies (which constitute the small, faint end
of the galaxy sample) lies at the heart of any weak lensing analysis.

\subsection{Weak lensing pipeline}

The unbiased measurement of galaxy shapes is not a trivial task,
because the observed images have been `corrupted': even in space based
data, the finite size of the mirror and the complicated telescope
optics give rise to a non-trivial point spread function (PSF). In
ground based data the situation is worse because of turbulence in the
atmosphere (an effect called seeing). Finally, the image is sampled in
discrete pixels (which may not be square), with a detector that may
suffer from charge transfer inefficiencies or other detector
non-linearities.

The combination of seeing and the intrinsic size of the PSF leads to a
circularization of the observed images, whereas PSF anisotropy
introduces coherent alignments in the shapes of the galaxies.  The
former effect lowers the amplitude of the inferred lensing signal; the
latter can 
mimic a lensing signal. Hence, to infer the true lensing signal, one
needs to determine the original galaxy shape: this requires some form
of deconvolution in the presence of noise. It is therefore not
surprising that the development of methods that can undo the effects
of the PSF has been a major focus of lensing research.

We list below the schematic steps of a pipeline that starts with raw
galaxy images and that ultimately delivers cosmological measurements (see
\cite{Jain06} for details and discussion of potential  
systematic errors at each step). 

{\bf 1. Object detection:} The detection of the faint galaxies that
are used in the analysis forms the first step in the lensing
analysis. This can be done on the individual exposures (as multiple
images of the same area of sky are typically obtained) or on a stacked
image. In either case, an algorithm to distinguish stars from galaxies
is needed.  Note also that the images need to be corrected for any
shearing by the camera. The next step is to quantify shape parameters
for these objects.  The optimal way to detect galaxies and measure
their shapes using multiple exposures and a set of filters is an area
of ongoing research which we will not address further.

{\bf 2. PSF estimation:} 
To deal with the effects of the point spread function (PSF) a sample
of moderately bright stars is identified from the actual data. These
are subsequently used to characterise the PSF in terms of its size,
second moment and possibly higher moments. 

The variation of the PSF across the field of view is described with
an interpolating function, which is typically done using a simple polynomial
model \cite{Hpsf}. This appears to be adequate for current results,
but ultimately the accuracy is limited by the fact that only a limited
number of stars is observed.  One way to address this issue is
to study the PSF in observations with a high number density of stars
\cite{Hpsf}. However, it is reasonable to assume that the PSF varies
in a relatively systematic fashion from exposure to exposure. This
allows one to decompose the observed patterns into their principal
components, as proposed by \cite{Jarvis04}. As the accuracy of cosmic
shear measurements improves, such a careful modeling of the PSF will
likely be required in order to undo the effects of the PSF.

{\bf 3. PSF correction:}
Having identified the galaxies of interest, the next step is to
correct the observed galaxy shapes for the convolution by the PSF.
This is arguably the most difficult, yet most important step in the
analysis. This is one of the most active areas of study and
innovation, not only to derive reliable results from the current
generation of surveys, but also to ensure that future, much more
demanding, cosmic shear surveys can reach their full potential.

A number of techniques have been developed to address this problem and
\cite{STEP1,STEP2} provide detailed descriptions of these. The most
widely used method, the KSB method, was developed by \cite{ksb}. This
is also one of the oldest methods, although modifications were
suggested by \cite{lk97,Hoekstra98}. The KSB method assumes that the
PSF can be described as the convolution of a compact anisotropic
kernel and a large isotropic kernel. Although this may appear to be a
reasonable assumption for ground based data, it is not correct for
space based data \cite{Hoekstra98}. A nice feature of the method is
that the correction for PSF anisotropy and the circularization by the
PSF are separate operations. The latter step (the closest to the
actual deconvolution) is typically performed by averaging the
correction for objects of similar size.

The KSB approach is limited by the assumptions that have to be made
about the PSF and galaxy profiles. Relaxing such assumptions and
expanding the object surface brightness distribution using a suitable
set of basis functions has attracted much attention in recent years
\cite{BJ02,Refregier03a,Refregier03c,Nakajima07}. A related approach is to fit
versatile models to the data and perform the deconvolution using the
best fit models. For instance \cite{Kuijken99} explored whether
objects can be modeled as sums of Gaussians. An advantage of the
model fitting methods is that pixelation effects are readily
implemented. Although progress is being made, it is currently
unclear what the best procedure will be.

{\bf 4. Measurement of shear correlations and cosmological parameters:}
Once a catalog of galaxy shapes is available, cosmological statistics
can be calculated. The two-point correlations of the shear are
calculated simply from the galaxy positions, ellipticities, and
weights that characterize the signal-to-noise of the shape
measurement. These may be calculated as a power spectrum or
correlation function for varying angular separations. For other
applications to galaxy and cluster lensing, one works with the
background shapes with reference to foreground objects. 

Cosmological inferences rely on one additional property of the source
galaxies, namely their redshift. The galaxies of interest are too
faint to be observed spectroscopically. Instead, weak lensing studies
rely on photometric redshifts (henceforth photo-z's). These are based
on the measured galaxy colors and other properties that provide a
coarse estimate of the galaxy's redshift. Improvements in photo-z
estimation and quantifying the effect of photo-z errors on
cosmological inferences from lensing are both active areas of
research.

\subsection{Diagnostics of the lensing signal}

An obvious concern is that one cannot ``see'' the weak lensing signal:
a weakly sheared galaxy appears unchanged, because of its much
larger intrinsic ellipticity. How can we be sure that the recovered
lensing signal is cosmological in nature, and not dominated by
observational distortions?

As discussed below, we can test the weak lensing analysis pipeline on
simulated data, but these simulations may lack a systematic effect
that is present in real data. Fortunately, a number of diagnostic
tools can be used to test the reliability of the recovered lensing
signal. These tests cannot guarantee whether the recovered signal is
free of systematics, but they do often indicate whether systematics are
present. (Note, however, that the correction for the circularization by
the PSF cannot be tested using the diagnostics discussed below.)

The first diagnostic makes use of the fact that the corrected galaxy
shapes should not correlate with the (uncorrected) shapes of stars
(e.g., \cite{Bacon03}). The measurement of this correlation
is sensitive to imperfections in the model for PSF anisotropy and
imperfections in the correction scheme itself. The former can be
tested by correlating the corrected shapes of stars \cite{Hpsf}.

Another unique diagnostic makes use of the fact that the weak lensing
shear arises from a gravitational potential. Consequently, the
resulting shear field is expected be curl-free (see, however,
\cite{Schneider02}). The observed ellipticity correlation functions
can be separated into two independent components, an ``E''-mode which
is curl-free and a ``B''-mode, which is sensitive to the curl of the
shear field \cite{Crittenden01,Schneider02}. Hence, the presence of a
significant ``B''-mode indicates that residual systematics
remain. (Note that the ``B''-mode may also have a physical origin, caused
by intrinsic alignments of the sources -- see \S4.2 for more
details.) 

A diagnostic of the cosmological nature of the lensing signal is its
variation with redshift. The distance factors in the lensing
efficiency function of Equation~\ref{eqn:weightgl} lead to a characteristic
variation with source redshift. Thus, provided the redshift estimates
are accurate, this variation can be used to test the cosmological
origin of the lensing signal. Another diagnostic that we discuss below is
the use of different statistical measures, such as two- and
three-point shear correlations, that have a distinct relationship for a
signal due to gravitational lensing. Finally, one can compare 
different lensing observables, such as those related to shear and to
magnification effects. 

\begin{figure}[t!]
\begin{center}
\includegraphics{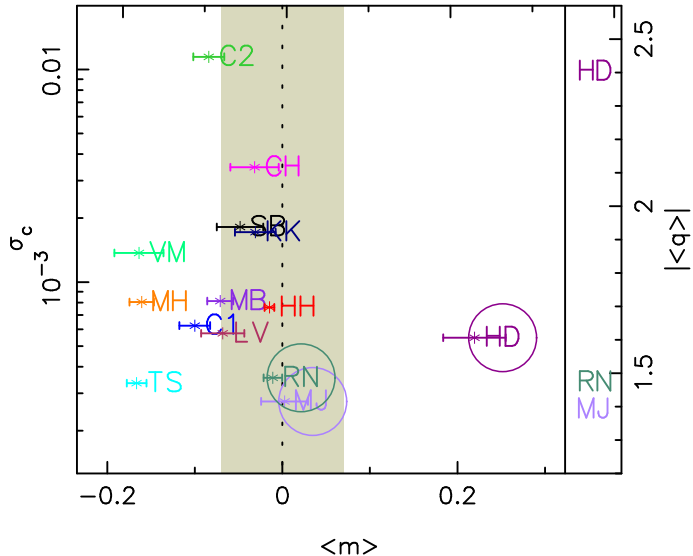}
\caption{\footnotesize Measurement of the the calibration bias $m$ and
PSF residuals $\sigma_c$ from \cite{STEP1}. The ideal method
has $m=0$ and small $\sigma_c$. The shaded region indicates
a bias of less than 7\%.  Methods that were used for the most recent
published cosmic shear results were found to have biases of
the order of a few percent. We refer the reader to \cite{STEP1}
for a detailed description of the symbols and methodology.}
\label{step}
\end{center}
\end{figure}

\subsection{Tests on simulated images}

The main hurdle in the shape measurement is a proper handling of the
PSF-induced systematics, not our lack of understanding of the relevant
physics. Our ability to correct for shape measurement systematics can
therefore be tested using simulated data, which is an important
advantage of lensing over other methods.

The Shear TEsting Programme (STEP) is a collaborative effort involving
much of the weak lensing community to improve the accuracy of weak
lensing measurements, in preparation for the next generation of cosmic
shear surveys.  The first STEP paper \cite{STEP1} involved the blind
analysis of simulated ground based images. The galaxies in this
simulation had relatively simple morphologies; however, despite these
limitations, the results provided an important benchmark for the
accuracy of current ground-based weak lensing analysis methods.

The results of this exercise are shown in Figure~\ref{step}.  The
dominant source of error is the correction for the size of the PSF,
which leads to an overestimate of the shear by a multiplicative factor
$(1+m)$. A blind analysis of more complicated galaxies was presented
in \cite{STEP2}, which also included an improvement in the statistical
accuracy of the test.  These two studies showed that pipelines that
have been used to constrain cosmological parameters can recover the
lensing signal with a precision better than 7\%, within the
statistical errors of current weak lensing analyses. The most
successful methods were shown to achieve 1-2\% level
accuracy. Although sufficient for current work, biases as a function
of object size and magnitude remain.  The next phase in this work is
to identify the points of failure and find improvements. The
simulations also need to become more realistic, for instance
through the inclusion of systematics at the detector level.



\section{Cosmic shear and dark energy}

\subsection{Two-point shear correlations and tomography}

To quantify the lensing signal, we measure the shear
correlation functions from galaxy shape catalogs. The
two-point correlation function of the shear, for source galaxies in
the $i-$th and $j-$th redshift bin, is defined as
\begin{equation}
\xi_{\gamma_i\gamma_j}(\theta)=\langle\bm{\gamma_i}(\bm{\theta}_1)
\cdot\bm{\gamma_j}^\ast(\bm{\theta}_2)\rangle.
\label{eqn:shearcorrelation}
\end{equation}
with $\theta = |\bm{\theta}_1 - \bm{\theta}_2|$. 
Note that the two-point function of the convergence is identical to
that of the shear. It is useful to
separate $\xi_\gamma$ into two separate correlation functions by 
using the $+/\times$ decomposition: 
the $+$ component is defined parallel or
perpendicular to the line connecting the two points
taken, while the $\times$ component is defined along $45^\circ$.  
This allows us to define the rotationally invariant two-point
correlations of the shear field:
$\xi_{+}(\theta)=\langle\gamma_{i+}(\bm{\theta}_1)
\gamma_{j+}(\bm{\theta}_2)\rangle$, and 
$\xi_{\times}(\theta)=\langle\gamma_{i\times}(\bm{\theta}_1)
\gamma_{j\times}(\bm{\theta}_2)\rangle.
$
The correlation function of Equation~\ref{eqn:shearcorrelation} is
simply given by $\xi_{\gamma_i\gamma_j} = \xi_+ + \xi_-$. 

The E/B mode decomposition discussed in \S2.3 is given by
linear superpositions of $\xi_{+}(\theta)$ and $\xi_{\times}(\theta)$
(though it involves integrals over all $\theta$). A more direct way to
perform the E/B decomposition is through the mass aperture variance,
$M_{\rm ap}^2(\theta)$, which is a weighted second moment of the
tangential shear measured in apertures. This provides a very useful
test of systematics in the measurements; we will not use it here, but
refer the reader to \cite{Schneider02}. All two-point statistics such
as $M_{\rm ap}^2(\theta)$ can be expressed in terms of the shear
correlation functions defined above. 

\begin{figure}[ht!]
\begin{center}
\includegraphics[width=8cm,angle=0,bb=100 -30 554 550]{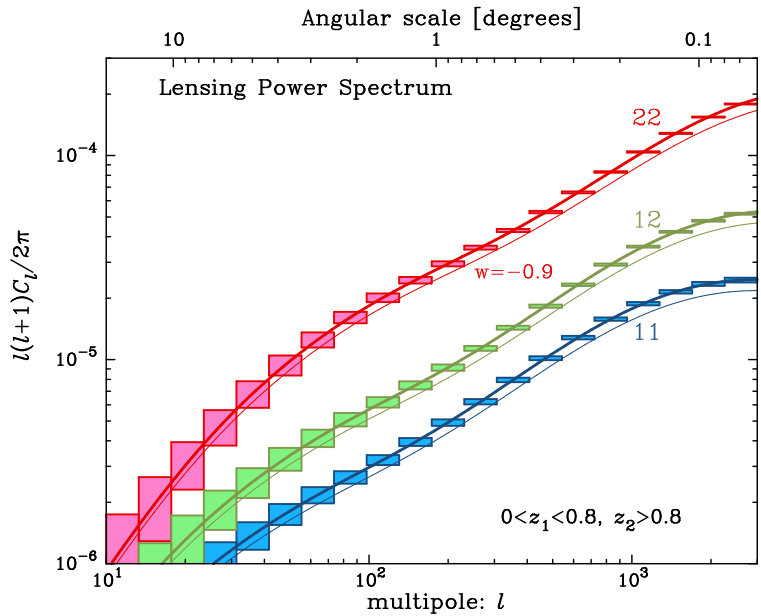}
\caption{\footnotesize 
The lensing power spectra constructed from
galaxies split into two broad redshift bins.  The two auto-spectra and one
cross-spectrum are shown. The
solid curves are predictions for the fiducial $\Lambda$CDM model, which
include nonlinear evolution \cite{Takada04}. 
The boxes show the expected measurement
error due to the sample variance and intrinsic ellipticity errors from
a 5000 deg$^2$ survey with median redshift $z=0.8$ (these are
ambitious survey parameters by the standards of Stage III surveys). 
The thin curves are the predictions for a dark energy model with $w=-0.9$. 
Note that at least four or five redshift bins are expected to be
useful from such a survey, leading to many more measured power spectra. 
\label{fig:pkappa}}
\end{center}
\end{figure}

The shear power spectrum at angular wavenumber $\ell$ 
is the Fourier transform of $\xi_{\gamma_i\gamma_j}(\theta)$. 
It is identical to the power spectrum of the 
convergence and can be expressed as a
projection of the mass density power spectrum $P_\delta$. 
For source galaxies in the $i$th and $j$th redshift bin it is
\cite{Kaiser92,Hu99} 
\begin{equation}
C_{\gamma_i\gamma_j}(\ell) = 
\int_0^{\infty} dz \,{W_i(z)\,W_j(z) \over \chi(z)^2\,H(z)}\,
 P_\delta\! \left ({\ell\over \chi(z)}, z\right ) .
\label{eqn:pkappa}
\end{equation} 
where the indices $i$ and $j$ cover all the redshift bins.  The
redshift binning is assumed to be provided by photo-z's that can be
estimated from multi-color imaging. If both source galaxy bins are
taken at redshift $z_s$, then the integral is dominated by the mass
fluctuations at a distance about half-way to the source galaxies.
Figure~\ref{fig:pkappa} shows the predicted auto and cross-spectra for
galaxies split into two redshift bins. The error bars show the sample
variance (which dominates at low $\ell$) and intrinsic ellipticity
(which dominates at high $\ell$) contribution to the measurement error. The
latter is also referred to as shape noise in the literature. (Note
that the measured power spectrum includes contributions from
systematic errors, which we discuss in \S4. )

%


Equation \ref{eqn:pkappa} shows how the observable shear-shear power
spectra are sensitive both to the geometric factors given by $W_i(z)$
and $W_j(z)$, as well as to the growth of structure contained in the
mass density power spectrum $P_\delta$. Both are sensitive to dark energy
and its possible evolution which determines the relative amplitudes of
the auto and cross-spectra shown in Figure~\ref{fig:pkappa}. $P_\delta$ also 
contains information about the primordial power
spectrum and other parameters such as neutrino masses. 
In modified gravity theories, the shape and time evolution of the density
power spectrum can differ from that of a dark energy model, even one
that has the same expansion history. Lensing is a powerful means of
testing for modifications of gravity as well 
\cite{Knox06,Amendola07,Jain07,Heavens07,Huterer07}. The
complementarity with other probes of each application of lensing is
critical,  especially with the CMB and with measurements of the
distance-redshift relation using Type Ia Supernovae and baryonic
acoustic oscillations in the galaxy power spectrum. 

The mass power spectrum is simply related to the linear 
growth factor $D(z)$ on large scales (low $\ell$):
$P_\delta \propto D^2(z)$. However, 
for source galaxies at redshifts of about 1, observable
scales $\ell \simgt 200$ 
receive significant contributions from nonlinear gravitational
clustering. So we must go beyond the linear regime 
using simulations or analytical
fitting formulae to describe the nonlinear mass power spectrum
\cite{Jain97,Jain00,White04,Francis07}. To
the extent that only gravity describes structures on scales larger
than the sizes of galaxy clusters, this can be done with high
accuracy. There is ongoing work to determine what this scale
precisely is and how to model the effect of baryonic gas on smaller
scales \cite{Zentner07}. 

\subsection{Cross-correlations and higher order statistics}

The cross-correlation of foreground galaxy positions with background
shear is also an observable. As discussed in \S6, it has been measured by
averaging the tangential component of the background galaxy
ellipticities in circular annuli centered on the foreground galaxy,
It is denoted $\langle\gamma_T\rangle(\theta)$ and 
is related to the Fourier transform of the galaxy-convergence power
spectrum, which in turn can be expressed analogously to the power
spectrum of equation \ref{eqn:pkappa}: 
\begin{equation}
C_{g_i\kappa_j}(\ell) = 
\int_0^{\infty} dz \,{W_{g i}(z)\,W_j(z) \over \chi(z)\,H(z)}\,
 P_{g\delta}\! \left ({\ell\over \chi(z)}, z\right ),
\label{eqn:pcross}
\end{equation} 
where $W_{gi}$ is the normalized redshift distribution of the
lens (foreground) galaxies and $P_{g\delta}$ is the three-dimensional
galaxy-mass density power spectrum. 

Along with the galaxy-galaxy power spectrum, $C_{g_i g_i}(\ell)$, 
equations \ref{eqn:pkappa} and \ref{eqn:pcross} represent
the three sets of auto- and cross-spectra that can be measured from
(foreground) galaxy positions and (background) galaxy shapes
\cite{Hu03}. Each of the three power spectra can be measured for multiple
photo-z bins. These contain all the two-point information one can
extract from multicolor imaging data on both galaxy clustering and
lensing. It would be an exhaustive exercise in parameter estimation to
perform model fitting on a set of such measurements; only pieces of this
have been carried out so far. 

In addition, cosmographic information via the 
distance-redshift relation can be obtained using the variation of the
galaxy-galaxy lensing signal with redshift. While this has less
constraining power than other tests of the distance-redshift relation,
it would help isolate the geometric and growth of structure information
that can be obtained from lensing
\cite{Jain03,Hu03,Bernstein04,Zhang05,Song04}. 

The combination of $C_{g_i g_i}(\ell)$ and $C_{g_i \kappa_j}(\ell)$ can
be used to determine the bias factor $b$ that relates the
three-dimensional galaxy power spectrum to that of the mass
density (the ratio of the two for appropriate redshift bins is
proportional to $b$). With this empirical determination of $b$,
cosmological parameters can be obtained more robustly from 
$C_{g_i g_i}(\ell)$ and its three-dimensional counterpart measured from
spectroscopic surveys \cite{Seljak05a}. Planned surveys with good
quality imaging and well calibrated photo-z's will enable cosmological
applications of $C_{g_i \kappa_j}(\ell)$ as well, once the bias is
known for the lens galaxies as a function of scale. 

Thus a variety of cosmological measurements can
be made from lensing and galaxy power spectra. The two main advances
awaiting these applications are reliable photo-z's and lensing shapes
measured over a wide area survey. 

Finally, the non-Gaussian properties of the lensing mass imprints
three-point and higher order correlations in the shear field. These
are valuable both for complementary cosmological information and for
checks on systematic errors \cite{Bernardeau97,Huterer06}. The 
lensing three-point function, or bispectrum in Fourier space, 
arises from nonlinear gravitational
evolution of the lensing mass. It vanishes at lower order in
perturbation theory, as the leading order (linear) density field is 
Gaussian random. The second order contributions to the bispectrum give
it special value: its dependence on cosmological parameters differs
from that of the power spectrum. In particular, it is possible to
combine the power spectrum and bispectrum to contrain the matter or
dark energy density with little dependence on the power spectrum
amplitude \cite{Bernardeau97,Jain97,Takada04}. 

The signal-to-noise ratio for a measurement of the bispectrum is lower than
for the power spectrum, and is more sensitive to the number density of
source galaxies. While three-point lensing correlations have been
detected in current data, it is expected to be useful for cosmology 
only in next generation datasets. Perhaps of equal importance is the
ability to test for systematic errors using three-point
correlations. \cite{Huterer06} showed how the degradation due to
systematics can be reduced by adding bispectrum and power spectrum
measurements. 

\subsection{Cosmological parameters}

\begin{figure}[t!]
\begin{center}
\includegraphics[width=7cm,angle=0,bb=100 0 554 450]{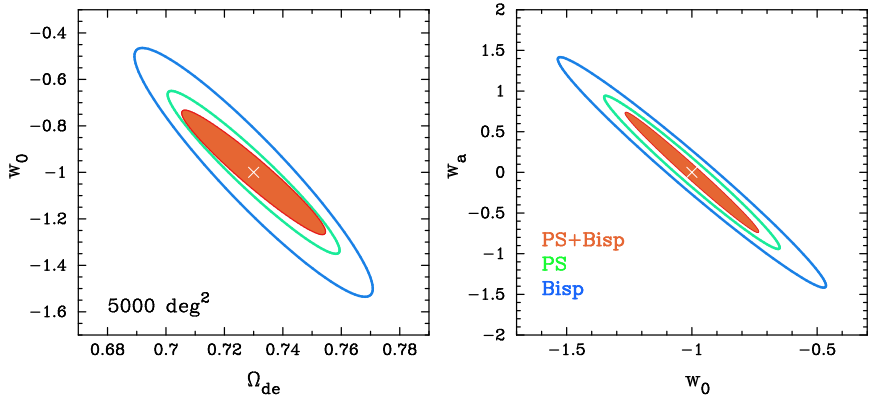}
\caption{\footnotesize 
Dark energy contours (68.3\% confidence level) 
from lensing power spectra and bispectra from an
ambitious Stage III survey, as in Figure~\ref{fig:pkappa}. 
The estimated non-Gaussian covariances between
the power spectra and bispectra are included in the joint
constraints. These forecasts assume Planck priors and do not include
systematic errors. 
\label{fig:contours}}
\end{center}
\end{figure}

Given a data vector, the Fisher information matrix describes how errors
propagate into the precision on cosmological parameters $p_\alpha$.
The Fisher matrix applied to the lensing power spectra is given by
\begin{equation}
F_{ij} = \sum_{\ell} \,
\left ({\partial {\bf C}\over \partial p_i}\right )^T\,
{\bf Cov}^{-1}\,
{\partial {\bf C}\over \partial p_j},\label{eq:latter_F}
\end{equation}
\noindent where ${\bf C}$ is the column matrix of the observed power
spectra and 
${\bf Cov}^{-1}$ is the inverse of the covariance matrix 
between the power spectra. 
The partial derivative with respect to a
parameter $p_\alpha$ is evaluated around the fiducial model. 
The Fisher matrix quantifies the best statistical errors achievable 
on parameter determination with a given data set:
the variance of an unbiased estimator of a parameter $p_\alpha$ obeys
the inequality: 
\begin{equation}
\langle\Delta p_\alpha^2\rangle\ge (\bm{F}^{-1})_{\alpha\alpha},
\end{equation}
where $(\bm{F}^{-1})$ denotes the inverse of the Fisher matrix and
$\Delta p_\alpha$ is the relative error on parameter $p_\alpha$ around
its fiducial value, 
including marginalization over the other parameters. 

This formalism has been used to forecast contraints on the dark energy
density $\Omega_{de}$ and equation of state parameters $w_0$ and
$w_a$. Lensing is also sensitive to other cosmological parameters that
affect either the primordial power spectrum or the growth of
structure \cite{HuTegmark,Song04,Takada04}. 
Since the projected power spectrum probed by lensing is a
slowly varying function of wavenumber (unlike the CMB), it is not
sensitive to parameters that produce localized features 
such as the baryon mass fraction. However its shape can help 
constrain neutrino masses and a running spectral index. Lensing
tomography is most sensitive to variables that affect the amplitude at
different redshift: this is what gives the greatest leverage on dark
energy parameters.  Thus the use of all the auto and cross-spectra that can be
measured with redshift information is critical in extracting 
cosmological information from lensing. 

Figure~\ref{fig:contours} shows dark energy forecasts obtained for a
5000 deg$^2$ Stage III experiment (in the terminology of the
Dark Energy Task Force \cite{Albrecht06}), 
with all other relevant cosmological parameters 
marginalized over \cite{Takada04}. CMB priors are used, at the level
of the Planck experiment.  Statistical errors due to sample variance
and finite intrinsic ellipticity are included, but the effect of systematic
errors is not: those are discussed and estimated in \S4
below. Aside from the issue of systematics, forecasts for lensing are
still challenging because there are multiple observables (power
spectra, bispectra, cluster counts). These observables are not
independent, and  much of the information is in the non-Gaussian
regime which makes  the estimation of covariances difficult. 
In Figure~\ref{fig:contours} we have used
power spectra and bispectra in three broad redshift bins, and included
covariances based on the halo model (Takada \& Jain, in preparation). 

\section{Systematic errors}

Weak lensing measurements are prone to a number of systematic
errors. The first category of systematics arise in the measurement of
galaxy shapes. But there are others that enter in the estimation of
cosmological parameters given a galaxy shape catalog. The primary
sources of systematic error can be characterized as: 

\begin{itemize}
\item{Knowledge of the PSF}
\item{Correction of the PSF and shear calibration}
\item{Intrinsic alignments}
\item{Photometric redshifts}
\item{Non-linear power spectrum/effect of baryons}
\end{itemize}

While each of these errors is well studied and can be modeled, small
residuals in the corrections of these errors may well be comparable to
statistical errors in lensing measurements.  The first two sources of
error follow directly from our discussion in \S2 of lensing shape
measurements. The remaining systematics become important if we wish to
interpret the lensing signal and compare it with a cosmological model.  

Intrinsic alignments are are caused by the tidal gravitational field, which
can cause the shape of a galaxy to be aligned with another due to
direct interactions. It is currently the least well characterized
source of systematic error in lensing. Photo-z
uncertainties can contribute to systematic errors because the
cosmological inferences of lensing measurements depend sensitively on
the estimated photo-z. Finally, the theoretical model predictions may
have uncertainties due to nonlinear gravitational clustering and
baryonic gas physics that affect the lensing power spectrum. Another
complication is the fact that we do not observe the shear directly,
but the reduced shear $\gamma/(1+\kappa)$ instead. We discuss the
systematic errors due to intrinsic alignments below, as these are the
least well characterized at present.

\subsection{Intrinsic alignments}

Thus far we have assumed that the galaxy ellipticities are
uncorrelated in the absence of lensing. However, there are reasons to
believe that this assumption is not valid and that intrinsic
alignments in the galaxy shapes contaminate the lensing signal. Two
kinds of intrinsic alignment effects have been identified. Before we
discuss these in more detail, we note that relatively little is known
about this effect, except that the alignments are relatively small.
It is safe to say that currently a robust measurement of this signal
is almost as difficult as that of the cosmic shear signal itself.

The first type to be identified is due to alignments of galaxy halos
with other halos that respond to tidal gravitational forces. Early
work based on simulations and analytic models \cite{Croft00,Crittenden01}
demonstrated how this process can compromise cosmic shear
measurements.  There is, however, considerable theoretical
uncertainity in modeling the alignment of halos themselves, and how
well the luminous matter (that is observed) aligns with the dark matter.
Importantly, this source of systematics can be greatly reduced with
photo-z's by using cross-spectra of galaxies in two different redshift
bins (so that galaxy pairs are separated by large distances at which
the tidal effects are very weak).

The second alignment effect was pointed out by \cite{Hirata04}. It
arises from the fact that the shapes of galaxies may be correlated
with their surrounding density field. This field is also
responsible for the weak lensing shear. As a result an
anti-correlation between the shapes of galaxies at different redshifts
is introduced, leading to a suppression of the lensing signal.  While
the first intrinsic alignment effect can be minimized by using
galaxies at different redshifts in shear correlation measurements,
this second mechanism affects pairs of galaxies at different
redshifts.  Without further theoretical and observational progress,
therefore, the two classes of intrinsic alignments can be a very difficult
systematic to overcome and can bias cosmic shear results (e.g.,
\cite{Bridle07}). Fortunately some handle on the level of these
effects is now available thanks to spectroscopic data
\cite{Mandelbaum06a,Hirata07} and theoretical progess via numerical
simulations is also being made \cite{Heymans06}. Nonetheless,
as shown by \cite{Bridle07}, intrinsic alignments need to be taken
into account when designing future surveys as they affect
the requirements on the accuracy of photo-z's.

\begin{figure}[t!]
\begin{center}
\includegraphics[width=6cm,angle=270,bb=0 400 654 450]{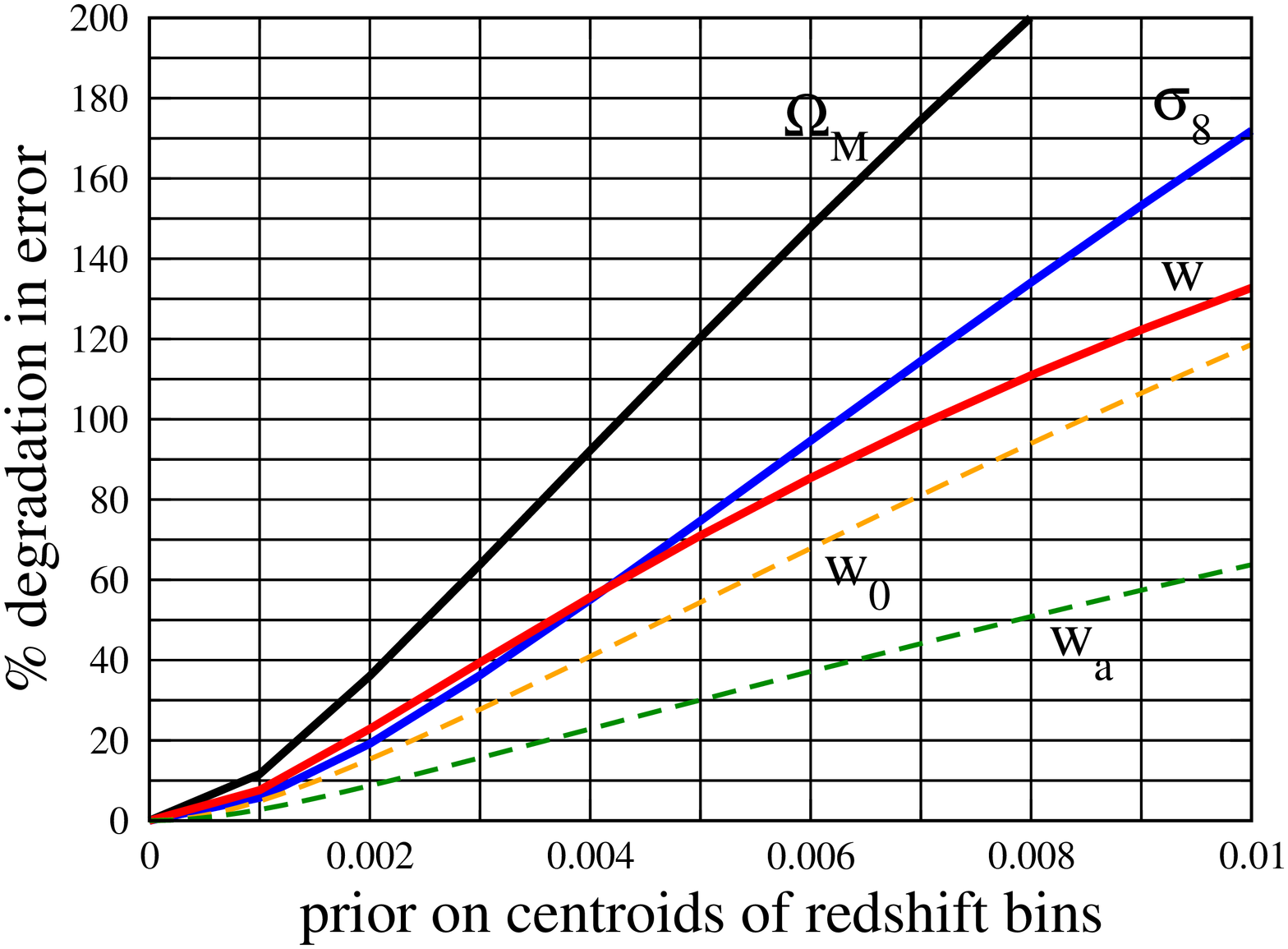}
\includegraphics[width=6cm,angle=270,bb=0 -300 654 450]{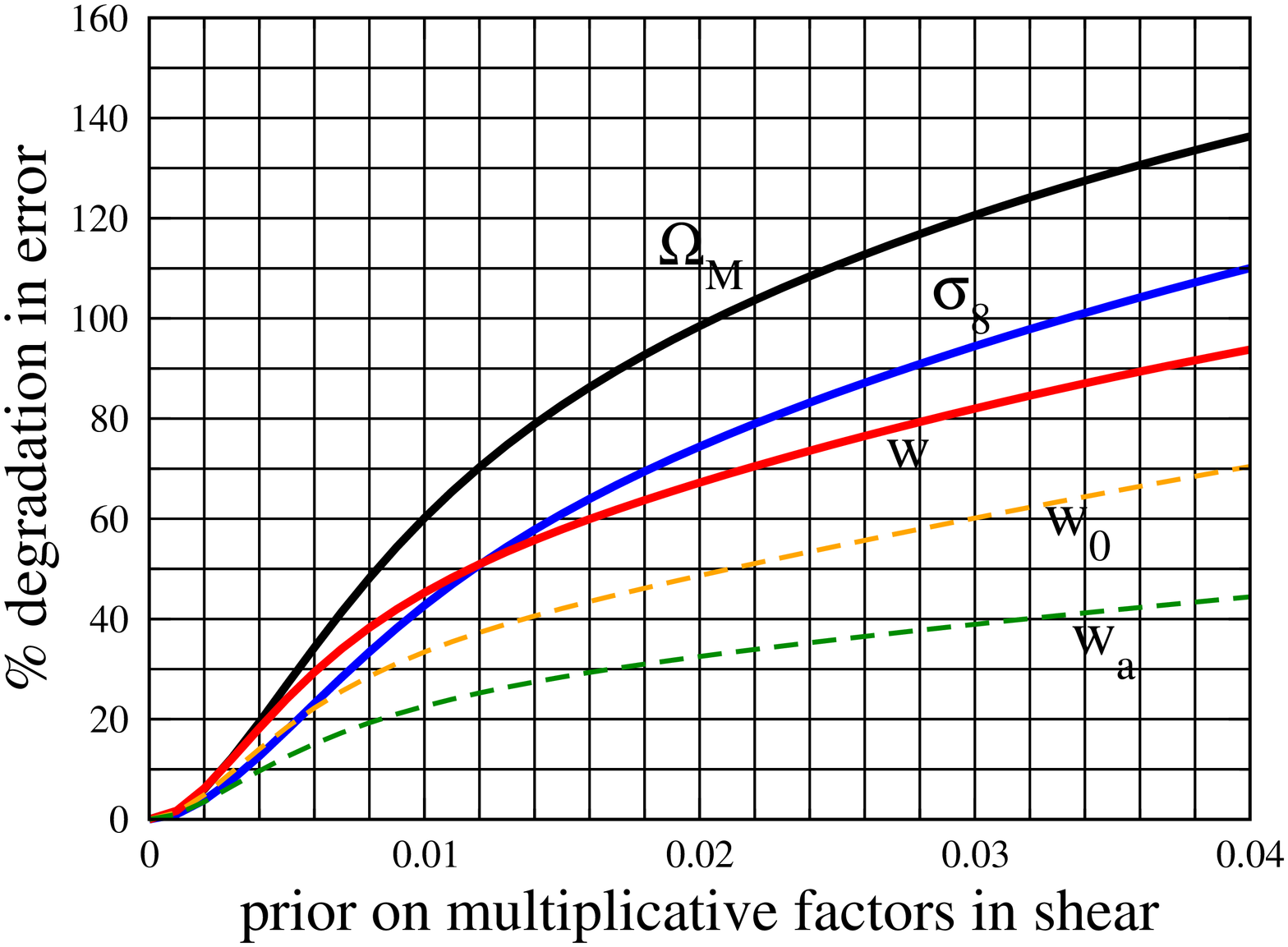}
\caption{\footnotesize 
Degradation in cosmological parameter accuracy due to systematic
errors \cite{Huterer06}.  
The left panel shows the effect of biases in photo-z's for a
Stage III survey.  The right panel shows the effect of shear
calibration errors. There is some evidence for a self-calibration
regime for these, unlike for photo-z biases. Note
that only the fractional degradation in parameter accuracy is shown here. 
\label{fig:degradation}}
\end{center}
\end{figure}

\subsection{How systematics degrade cosmological constraints}

A comprehensive study of lensing systematics can be made with the
following general expression of the estimated shear \cite{Huterer06}: 
\begin{equation}
\hat{\gamma}(z_s, {\bf n}) = \gamma_{lens}\left (z_s, {\bf n}\right )
    [1+\gamma_{sys}^{mult}(z_s, {\bf n})] \ + \ \gamma_{sys}^{add}(z_s,{\bf n})
\end{equation}
The above equation includes two kinds of systematic error
contribtutions, which modify the lensing shear via 
additive and multiplicative terms. In addition, the bin redshift $z_s$
and its width may also be in error, 
leading to biases in cosmological parameters \cite{Ma06}. 

Two key points in understanding the degradation of cosmological
information due to systematics are: (i) The impact of systematics on
shear correlations is what matters, e.g. errors that affect individual
galaxy shapes but are uncorrelated between galaxy pairs simply act as
additional statistical errors (and are likely to be subdominant to the
intrinsic shape noise of galaxies). (ii) Systematic errors typically
do not share the full redshift dependence of the lensing signal. For
example, errors in shear calibration may depend on galaxy size and
brightness, but not directly on redshift. This allows us to fit for
uncertainty in the shear calibration from the data. In general, by
using all available auto- and cross-spectra, one can 
marginalize over a set of systematic error parameters. 
(The exception is photo-z biases, which 
can mimic the cosmological signal and must be controlled with an
appropriate calibration sample.)

Figure~\ref{fig:degradation} shows how two of these systematic errors
degrade cosmological parameter accuracy: shear calibration errors and
photo-z biases. The right panel in this figure shows
the possibility of self-calibration: the degradation plateaus
somewhat, because the redshift dependence of the
lensing signal allows for joint measurements of parameters describing
both systematics (shear calibration in this case) and cosmology.  The
modeling and reduction  of systematics is an area of active study
(e.g. \cite{amara07}). The degradation estimates shown in
Figure~\ref{fig:degradation}  are meant to be conservative in that no
assumptions are made about the functional forms of these systematics
and only the shear power spectra are used. 

In summary, progress in handling systematic errors is being made 
on several fronts: for PSF effects, in the use of improved algorithms;
for photo-z's, in the use of large calibration samples; for intrinsic
alignments, via new measurements as well as physical models; and for
predictions in the nonlinear regime, from N-body and hydrodynamical 
simulations. The scientific returns of planned surveys
rely on continued advances in these directions. 

\section{Observational results}

The advent of wide field imaging cameras on 4m class telescopes in the
late 1990s made the first cosmic shear detections possible
\cite{bacon00,kaiser00,vw00,wittman00}. Since then, the size of weak
lensing surveys has increased significantly, and current surveys have
imaged several tens of square degrees of the sky. An extensive list of
early work can be found in \cite{Refregier03b}, but here we limit the
discussion to the relatively large surveys listed in
Table~\ref{survey}. We only 
include references to the most up-to-date analyses, which include a
separation of the signal into `E' and `B'-modes. The results listed in
Table~\ref{survey} all find small or negligible `B'-modes, in
particular on scales larger than a few arcminutes. Furthermore, the
lensing pipelines used to obtain these measurements are amongst the
most accurate ones available (see \S2.4 and \cite{STEP1,STEP2}). The
cosmological measurements from these surveys are discussed below in
\S5.1.   

\begin{table*}[t!]
\caption{Overview of recent surveys. The survey size and inferred
  value of $\sigma_8$ and $w$ (where available) 
are shown for the large ground based surveys and
  the largest spaced based survey. The $\sigma_8$ values are quoted at
  fixed $\Omega_m$ to check the consistency of the measured lensing
  amplitude in different surveys. Meaningful confidence intervals 
  on $\sigma_8$ require a joint
  analysis with CMB data (see text in \S5.1). Further, since these
studies have dealt with  marginalization and systematic errors in
  different ways, direct comparisons are difficult.} 
\label{survey}
\begin{tabular}{lll|lll}
\toprule
Survey & Area [deg$^2$] & Ref. & $\sigma_8$ ($\Omega_m=0.3$) & $w$ & Ref.\\
\colrule
RCS        & 53  & \cite{Hoekstra02b}    & $0.65\pm0.07$ &             & \cite{Benjamin07}   \\
VIRMOS     & 8.5 & \cite{vw05}           & $0.83\pm0.06$ &             & \cite{Benjamin07}   \\
CTIO$^a$       & 70  & \cite{Jarvis06}       & $0.81^{+0.15}_{-0.10}$  & $-0.89^{+0.16}_{-0.21}$ & \cite{Jarvis06}\\
GaBoDS     &  13 & \cite{Hetterscheidt07}& $0.78\pm0.08$ &             & \cite{Benjamin07}        \\
\colrule
CFHTLS$^{b,c}$ & 57 & \cite{Fu07}        & $0.71\pm0.04$ & $<-0.5$  & \cite{Fu07}     \\
COSMOS     & 2   & \cite{Massey07a}      & $0.87^{+0.09}_{-0.7}$ &  & \cite{Massey07a}\\
\botrule
\end{tabular}
\begin{footnotesize}
\\$^a$: error bars correspond to 95\% confidence limits and are joint
constraints using WMAP priors; $^b$: published analyses do not include
full area and are based on $i'$ data only. The completed survey will
image 140 square degrees in 5 filters; $^c$: the upper limit on $w$ is
from \cite{Hoekstra06}
\end{footnotesize}
\end{table*}

Constraints on cosmological parameters have not improved dramatically,
because of the limited knowledge of the source redshifts. The ability
to measure the growth of structure as a function of redshift will be
the next major leap forward for cosmic shear studies. The first steps
in that direction have already been taken and the bottom three surveys
in Table~\ref{survey} are able to derive photometric redshifts for the
sources thanks to their multi-color data\footnote{A fourth survey, the
Deep Lens Survey, has also collected multi-color data, but no recent
results have been published}. The first tomographic results have
already been presented\cite{Kitching07,Massey07a}, 
albeit based on small areas which leave them
susceptible to non-Gaussian contributions to the sample variance. 
The COSMOS survey combines the excellent
image quality of the Hubble Space Telescope with an extensive
follow-up campaign to map the three-dimensional large scale
structure \cite{Massey07b} and measure the cosmic shear signal as a
function of redshift \cite{Massey07a}.  Expanding space based
observations beyond the COSMOS survey area will require a dedicated
space based mission (see \S7); ground based surveys will lead
the way in the near future.

\begin{figure}[ht!]
\begin{center}
\includegraphics[width=10cm,angle=-90]{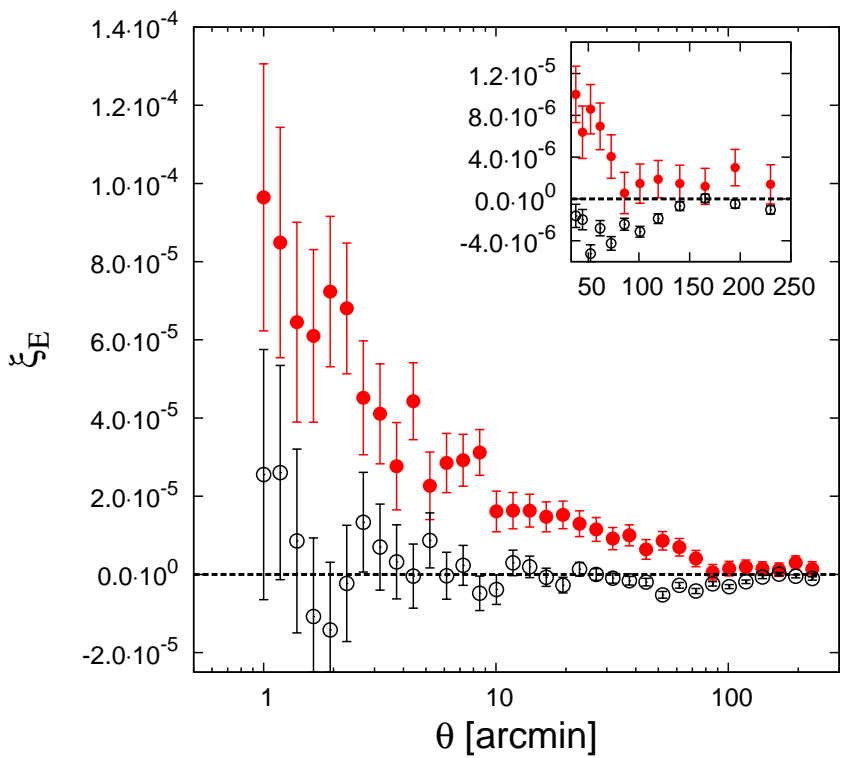}
\caption{\footnotesize Ellipticity correlation function from \cite{Fu07}. These
measurements based on the analysis of 57 deg$^2$ of CFHTLS $i'$
imaging data, extend out to 4 degrees, well into the linear regime. The
E-modes are indicated by the red points. The B-mode (open points) is
consistent with zero on most scales. As shown in the enlargement,
there is an indication of residual systematics on a scale of one
degree, which corresponds to the size of the camera.
\label{cfhtls}}
\end{center}
\end{figure}

\paragraph{Canada-France-Hawaii-Telescope Legacy Survey:}

The CFHTLS is the largest cosmic shear survey
carried out to date. Once completed, by the beginning of 2009, it will
have imaged 140 square degrees in the five Sloan filters, and galaxies
as faint as $i'=24.5$ are included in the analysis. The availability
of photometric redshifts for the sources will improve the constraints
on cosmological parameters significantly. The goal of the completed
survey is to constrain $w$ with a relative accuracy of $5-10\%$.

The first results, based on a conservative analysis of the
first year $i'$ data were presented in \cite{Hoekstra06}, whereas
\cite{Semboloni06} measured the signal from the ``deep'' component of
the survey. More recently, \cite{Fu07} analysed 57 deg$^2$ of CFHTLS
$i'$. As shown in Figure~\ref{cfhtls}, \cite{Fu07} were able to
measure the lensing signal out to 4 degrees, well into the linear
regime.

\subsection{Implications for cosmology}

As discussed above, our 
limited knowledge of source redshifts is still a significant source
of uncertainty in many lensing studies. This situation is improving rapidly,
however, thanks to multi-color surveys such as COMBO-17, COSMOS and
CFHTLS. This has led to an updated analysis of the RCS
\cite{Hoekstra02b}, VIRMOS \cite{vw05}, the Garching-Bonn Deep Survey
(GaBoDS;\cite{Hetterscheidt07}) and first year CFHTLS \cite{Hoekstra06}
measurements by \cite{Benjamin07}. Compared to the original
measurements, \cite{Benjamin07} include a more sophisticated treatment
of the sample variance errors on small scales, as suggested by
\cite{Semboloni07}. The new analyses also include, where necessary,
corrections to the signal based on the STEP results
\cite{STEP1,STEP2}. Most importantly, they use up-to-date redshift
distributions for the surveys based on the large photometric redshift
catalog published by \cite{Ilbert06}.

\begin{figure}[ht!]
\begin{center}
\includegraphics[width=10cm]{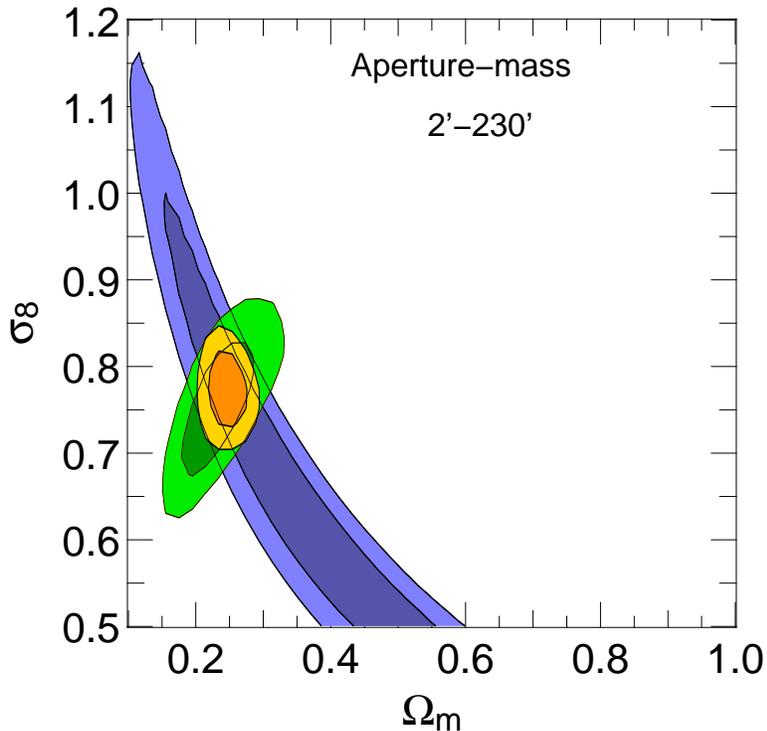}
\caption{\footnotesize Joint constraints on $\Omega_m$ and $\sigma_8$
from the CFHTLS \cite{Fu07} (purple) and WMAP3 \cite{Spergel07}
(green). The CFHTLS results are based on the aperture mass statistic
on scales ranging from $2'-230'$. The combined constraints from
weak lensing and CMB are indicated by the orange region, demonstrating
excellent agreement. Also note the complementarity of lensing to
CMB observations.\label{cosmo}}
\end{center}
\end{figure}

Due to the lack of tomographic measurements most lensing results
only constrain a combination of $\Omega_m$ and $\sigma_8$. This is
demonstrated clearly in Figure~\ref{cosmo}, which shows in purple the
weak lensing results from \cite{Fu07}. The various estimates for
$\sigma_8$ are listed in Table~\ref{survey} (adopting a $\Lambda$CDM
cosmology with $\Omega_m=0.3$ for reference, except for
\cite{Jarvis06} who have marginalized over all parameters in
combination with CMB data).
The ensemble averaged value for $\sigma_8$ (adopting $\Omega_m=0.3$)
from these recent measurements is $\sigma_8=0.75\pm0.03$ with a
$\chi^2=7.4$ with 5 degrees of freedom. The agreement between these
results is therefore reasonable (the probability of a larger $\chi^2$
is 0.19). We note, however, that this statistical error ignores any
common systematics. 

As shown in Figure~\ref{cosmo} the combination of lensing and CMB
measurements is useful in constraining $\Omega_m$ and $\sigma_8$. This
joint analysis from \cite{Fu07} with the WMAP3 results
\cite{Spergel07} yields $\Omega_m=0.248\pm0.019$ and
$\sigma_8=0.771\pm0.029$. These are fully consistent with the previous
lensing based analysis from the CTIO survey of \cite{Jarvis06}. 
These results are also in agreement with recent studies of the number
density of clusters of galaxies (e.g., \cite{Henry04,Rosati02}; also
see \cite{Hetterscheidt07} for a compilation of recent measurements).
Measurements of $w$, the dark energy
equation of state, are still limited due to the lack of tomographic
results from large area surveys, but two tentative results from analyses
with constant $w$ are listed in Table~\ref{survey}.

\section{Lensing by galaxies and galaxy clusters}

The study of cosmic shear has been the main science driver of most
recent weak lensing studies, but in this section we highlight
some of the applications to galaxy and cluster lensing 
which pertain to cosmology and the study of dark matter. 

\subsection{Mapping the distribution of dark matter}

The observed weak lensing shear field provides estimates of the
derivatives of the lensing potential (see Eqn.~\ref{sheardef}).  As
shown by \cite{KS93} it is possible to invert this problem to
obtain a parameter-free reconstruction of the surface density
distribution: it is possible to
make an `image' of the dark matter distribution. The surface density
(up to an arbitrary contant $\kappa_0$) can be written as \cite{KS93}:

\begin{equation}
\kappa({\bm\theta})-\kappa_0=\frac{1}{\pi}
\int d^2{\bm \theta'}
\frac{\zeta({\bm \theta'-\theta}) \gamma({\bm \theta'})}
{({\bm \theta'-\theta})^2},
\end{equation}

\noindent where the convolution kernel $\zeta({\bm \theta})$ is 
given by

\begin{equation}
\zeta({\bm \theta})=\frac{\theta_2^2-\theta_1^2+2i\theta_1\theta_2}
{|{\bm \theta}|^4}.
\end{equation}

\noindent The proper evaluation of this integral requires data out to
infinity, which is impractical. This complication spurred the
development of finite-field inversion methods
\cite{Seitz96,Seitz01,Squires96}.

\begin{figure}[t!]
\begin{center}
\epsfxsize=\hsize
\epsffile{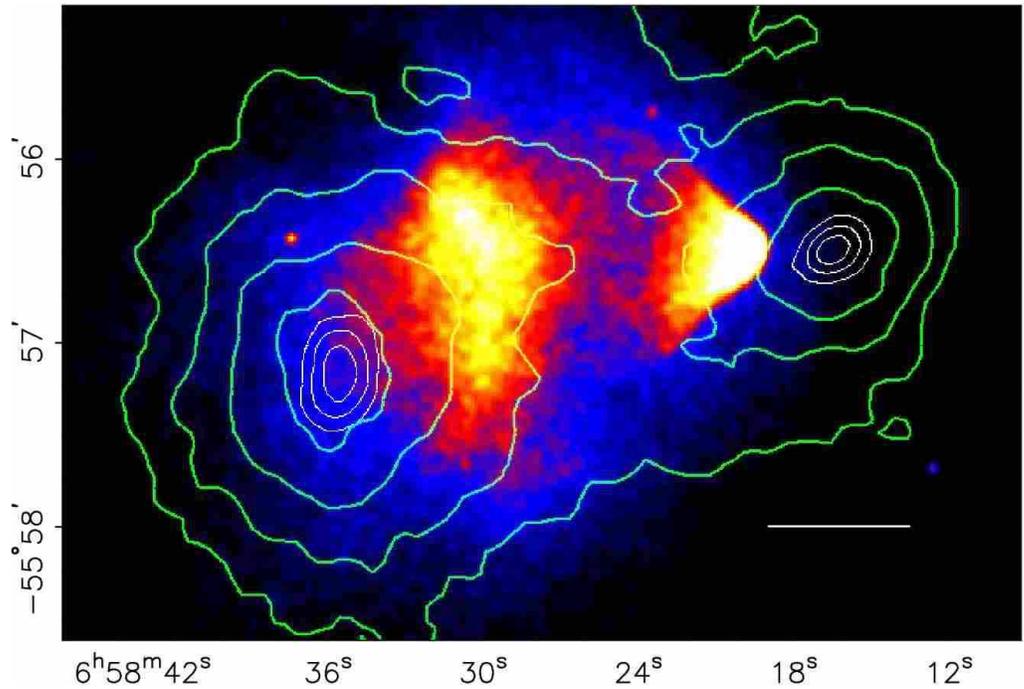}
\caption{\footnotesize X-ray emission from the `Bullet' cluster of
galaxies as observed by Chandra. The eponymous bullet is a small
galaxy cluster which has passed through the larger cluster and whose
hot gas is seen in X-rays as the triangular shape on the right.
The contours correspond to the mass reconstruction
from \cite{Clowe06}. The dark matter distribution is clearly offset
from the gas, which contains the majority of baryonic matter, 
but agrees well with the distribution of galaxies -- as expected if both
the dark matter and stars in galaxies are effectively collisionless. 
See \cite{Clowe06} for a complete discussion of this intriguing  object.
\label{bullet}}
\end{center}
\end{figure}

The intrinsic shapes of the sources add significant noise to the
reconstruction and as a result only the distribution of matter in
massive clusters of galaxies can be studied in detail using weak
lensing mass reconstructions. Of particular interest is the study
of merging systems, where dynamical techniques cannot be used
\cite{Clowe06,Jee07,Mahdavi07}.  Figure~\ref{bullet} shows a
reconstruction of the mass distribution of the Bullet cluster by
\cite{Clowe06} based on HST observations. The reconstructed (dark)
matter distribution is offset from the hot X-ray gas, but agrees well
with the distribution of galaxies. As shown by \cite{Clowe06}, these
observations provide some of the best evidence for the existence of
dark matter (also see the discussion in\cite{Angus07}). This is
because in alternative theories of gravity (such as Modified Newtonian
Dynamics \cite{Milgrom83,Bekenstein04}) the hot X-ray gas should be
the main source of the lensing signal. In the near future we can
expect improved constraints on the properties of dark matter particles
based on a systematic study of merging systems. 

\subsection{Cosmology with galaxy clusters}

Since clusters trace the highest peaks in the density, their number
density as a function of mass and redshift depends strongly on the
underlying cosmology (e.g., \cite{Rosati02,Bahcall03}), making it an
interesting complementary probe for dark energy studies
(e.g.,\cite{levine02}). Although conceptually straightforward, the
implementation of this method has proven difficult.

One reason is that precise measurements of cosmological parameters
require cluster catalogs with well-defined selection functions. In
principle clusters can be identified in mass reconstructions from
large weak lensing surveys (e.g., \cite{wittman06,miyazaki02,
miyazaki07}), but projections along the line-of-sight lead to a
relatively high false positive rate (e.g.,
\cite{White02,Hennawi05}). Hence, either one must work with a
statistic that includes projection effects (e.g. \cite{Marian06}), or
with samples derived from optical, X-ray or radio observations.

Even in the latter case it is essential to have a
well-determined relation between the observed cluster properties and
mass. This is where weak lensing studies of large cluster samples can
play an important role.  The determination of the mean relation
between the quantity of interest (e.g., richness, X-ray temperature)
and cluster mass can be done statistically. For instance,
\cite{Sheldon07,Johnston07} have measured the ensemble averaged weak
lensing signal as a function of richness and luminosity using data
from Sloan Digital Sky Survey (SDSS). 
Unfortunately, the mass-observable is expected to have an
intrinsic scatter as well, which is the result of differences in formation
history, etc. The precise characterization of this unknown scatter is
important to ensure accurate measurements of cosmological
parameters. Individual weak lensing masses can be derived for massive
clusters. We note, however, that ultimately the accuracy of these
mass measurements is limited by projections along the line of sight
\cite{Metzler99,Metzler01,Hoekstra01,White02}.

Multi-wavelength observations of samples that contain up to $\sim 50$
massive clusters have only recently started \cite{Dahle02,Hoekstra07,
Bardeau07}. These comprehensive studies, which also combine data at
other wavelengths, will not only help quantify the scatter, but will
also improve our understanding of cluster physics. This in turn will
increase the reliability of other cluster mass estimators (such as the
X-ray temperature). For instance, recently \cite{Mahdavi08}, found
evidence that the outer regions of clusters are not in hydrostatic
equilibrium, suggesting that additional pressure may be provided by
bulk motion of the plasma \cite{Nagai07}. Cluster cosmology is an
evolving field, and the hope is that with large samples of clusters
observed in multiple wavelengths their internal physics will be
modeled well enough for cosmological applications.

\subsection{Properties of dark matter halos}

Simulations of hierarchical structure formation in CDM cosmologies
have shown that the density profiles of virialized halos over a wide
range in mass have a 
nearly universal profile with radius -- the Navarro-Frenk-White (NFW) profile
\cite{nfw96,nfw97}.  The only difference between halos of galaxies and
clusters of different mass is their concentration, which reflects the central
density of the halo.  Gravitational lensing provides us with powerful
tools to test a range of predictions of the CDM paradigm via the
structure of halos. For instance, the dark matter dominated outer
regions can be uniquely probed by weak lensing, whereas strong
gravitational lensing can be used to study the density profile on small scales.

\paragraph{Central regions:}

In the context of CDM, simulations indicate a power law density
profile $\rho\propto r^{-\beta}$ as $r\rightarrow 0$. The original studies
\cite{nfw96,nfw97} found a slope of $\beta=1$, but the exact value is
still debated \cite{Moore99, Navarro04} . Without a complete treatment
of the effects of baryons, observational results will be difficult to
interpret. Despite these complications, much effort is devoted to
determine the slope of the density profile observationally, as it can
provide unique constraints on physical properties of the dark matter
particle, such as its interaction cross section
(e.g. \cite{Spergel00,Meneghetti01}).

Dynamical studies of galaxies have proven useful, and much of the
current controversy about the central slope is based on observations
of the rotation curves of low surface brightness galaxies, which
suggest that the dark matter distribution has a central core
(e.g.,\cite{Flores94,Moore94,McGaugh98}). Strong lensing by galaxies
can provide limited information because the typical Einstein radius is
large compared to the region of interest. Nevertheless, the
combination of strong lensing and dynamics has shown to be extremely
useful for the study of the stars and dark matter in galaxies (e.g.,
\cite{Treu02,Koopmans06}) and to test general relativity \cite{Bolton06}.

Strong lensing can be used to study the inner density profiles of
clusters, although results are still somewhat ambiguous
\cite{Comerford06}. Of particular interest are clusters that show both
tangential and radial arcs, because these can help to constrain the
density profile. An analysis of such systems by \cite{Sand04} suggests
an average slope $\beta\sim0.5$. However, \cite{Meneghetti07} studied
simulated clusters and found that too restrictive assumptions can bias
core slope estimates to lower values (also see \cite{Dalal03,Bartelmann04}).

\paragraph{Outer regions:} 

The value for the outer slope of the density profile is expected to be
$\beta\sim 3$.  A related prediction is that the mean central density
of the halo decreases with virial mass, i.e., lower mass systems are
more concentrated \cite{nfw96,nfw97}. The average dark matter profile
of galaxy clusters has been studied by \cite{Johnston07} using SDSS. These
results and measurements by \cite{Mandelbaum06d,Comerford07}
agree well with predictions from $\Lambda$CDM models, as do studies of
individual clusters such as Abell 1689\cite{Broadhurst05}.

The study of the outer parts of galaxies is more difficult, because
the signal of an individual galaxy is too small to be
detected. The interpretation of the observed signal, also
known as the galaxy-mass cross-correlation function (e.g.,
\cite{Hoekstra04, Sheldon04,Seljak05a}) is complicated by the fact
that it is the convolution of the galaxy dark matter profile and the
(clustered) distribution of galaxies. Despite these limitations,
galaxy-galaxy lensing studies provide a number of useful tests of the
cold dark matter paradigm.

One such test is the measurement of the extent of dark matter halos.
Pioneering studies by \cite{Brainerd96,Hudson98} were unable to
provide constraints because of the small numbers of lens-source
pairs. Large surveys, such as SDSS (e.g., \cite{Fischer00, Sheldon04,
Mandelbaum06b}), RCS \cite{Hoekstra04} and CFHTLS \cite{Parker07} have
measured the lensing signal with much higher precision, enabling
\cite{Hoekstra04} to determine the extent of dak matter halos
around field galaxies. Note that these measurements use the small
scale end of the galaxy-shear cross-correlation discussed above in
\S3.2. 

Another area where galaxy-galaxy lensing studies 
will have a great impact is the study
of the shapes of dark matter halos. CDM simulations predict that halos
are tri-axial (e.g \cite{Dubinski91}). This is
supported by the findings from \cite{Hoekstra04}, who found that the
dark matter halos are on average aligned with the light distribution
with a mean axis ratio that is in broad agreement with the CDM
predictions. A similar result was obtained recently by \cite{Parker07}
using CFHTLS data. Both these studies lacked the multi-color data to
separate lenses by galaxy type. Such a separation was done by
\cite{Mandelbaum06c} using SDSS data. They did not
detect a significant flattening, although their data do suggest a
positive alignment for the brightest ellipticals. 

The accuracy of these measurements is expected to improve
significantly over the next few years as more data is collected
as part of cosmic shear surveys.  An accurate measurement of the
anisotropy of the lensing signal around galaxies (i.e, the signal of
flattened halos in CDM) is also a powerful way to test alternative
theories of gravity \cite{Mortlock01,Hoekstra04}.

\section{The Future}

The first cosmic shear results, published less than a decade ago, were
based on areas of several square degrees at most
\cite{bacon00,kaiser00,vw00,wittman00}. Current leading surveys, most
notably the CFHTLS, are many times larger and provide photometric
redshift information for the sources. Even so, the area coverage of
CFHTLS is still modest (below 200 deg$^2$) given that it is now
technologically feasible to image more than a thousand square degrees
per year to depths of interest for cosmology. Furthermore,
despite the recent success in measuring the cosmic shear signal,
current data are obtained using telescopes that are not optimized for
weak lensing. Finally, the use of photo-z's for cosmological
inferences is still in its early stages. The CFHTLS lacks good
coverage at near-infrared wavelengths, which impacts the accuracy of
photo-z's.

\subsection{Planned surveys}

The next generation of surveys 
will address some or even all of these limitations. For instance, many
will make use of new telescopes specifically designed to provide a
stable PSF with minimal anisotropy. Equipped with new cameras with
fields of view of a square degree or larger, these surveys will
deliver over 1000 deg$^2$ of well calibrated images of
galaxies beyond $z=1$.  Table~\ref{future} lists some basic information for
a number of these projects. As many of these surveys are still in the
planning stage, we stress that these numbers may change. Experience
shows that this is particularly true for the starting dates! What is
clear from this table is that the data size will increase by another
order of magnitude in approximately the next five years.

Of the first four surveys listed in Table~\ref{future}, the Kilo
Degree Survey\footnote{http://www.strw.leidenuniv.nl/$\sim$kuijken/KIDS/}
(KiDS) and the Panoramic Survey Telescope and Rapid Response
System\footnote{http://pan-starrs.ifa.hawaii.edu} (PanSTARRS) use
telescopes and cameras specifically designed for the projects. The
Dark Energy Survey\footnote{http://www.darkenergysurvey.org} (DES)
will use a new camera built for the 4m Blanco telescope at
CTIO. Similarly, a new camera, the HyperSuprimeCam, has been proposed
for the Subaru 8.2m telescope. KiDS will use near-infrared imaging
with the VISTA telescope; the resulting 9-band data will give it
excellent photometric redshift accuracy, which is critical for the
measurement of intrinsic alignments and cosmic shear tomography.  DES
also plans to use near-IR imaging over a substantial part of the survey.

\begin{table*}[t!]
\caption{Overview of planned surveys}
\label{future}
\begin{tabular}{llllll}
\toprule
Survey          & Start & Area      & $n_{\rm eff}$    & Ground/Space \\ 
              &       & [deg$^2$] & [arcmin$^{-2}$] & \\
\colrule
KiDS          & $>2008$ & 1500    & $\sim 10$ & ground \\
PanSTARRS$^a$ & $>2008$ & 30,000  & $\sim 4$   & ground \\
DES           & $>2010$ & 5000    & $\sim 10$ & ground \\
Subaru        & $>2012$ & 2000    & $\sim$ 20-30 & ground \\
\colrule
LSST          & $>2014$ & 20,000  & $\sim$ 30-40 & ground \\         
SNAP          & $>2015$ & 4000    & $\sim 100$    & space  \\
DUNE          & $>2015$ & 20,000  & $\sim 40$     & space  \\
\botrule
\end{tabular}
\begin{footnotesize}
\\$^a$ Here we consider PS1, the PanSTARRS project with a single
telescope and the $3\pi$ survey. We note that this project may be
expanded into a four telescope project in subsequent years (referred
to as PS4).

\end{footnotesize}
\end{table*}

The accuracy with which the lensing signal can be measured depends on 
the area covered and the number density of distant source galaxies 
for which reliable shapes and photo-z's can be determined. For a fixed
amount of observing time, one therefore has to strike a balance between the
amount of sky covered and the depth of the observations. If the survey
is too shallow, most sources will be at low redshift, and the induced
lensing signal will be too small to be of interest.

On the other hand, increasing the number density of sources used in
the analysis is useful only if systematic effects for these small,
faint galaxies can be controlled or corrected for.  The blurring of
the images by the atmosphere is a limiting factor. If the size of the
galaxy is comparable to that of the PSF, any residual systematic is
amplified by the ``deconvolution''. Compared to space based
observations, ground based observations are therefore more sensitive
to problems with PSF correction. Furthermore, obtaining deep near
infrared (NIR) data, needed for accurate photometric redshifts, will
be more difficult for a ground based telescope (though it would be feasible
to add imaging in these bands from space provided the depth and sky
coverage match). 

Much of the cosmological information in planned surveys is extracted
from large scale modes, for which sample variance is the limiting
factor, not the number density. Consequently, ground based surveys may
suffer only a modest loss of accuracy if only galaxies with well
measured shapes and photo-z's are used. Further, the cost of a space
based project requires a careful consideration of the benefits. While
many of these considerations are being worked on, it is clear that the
requirements to reach percent level accuracy in the dark energy
equation of state are very challenging.

In Table~\ref{future} we 
provide crude estimates for the expected effective source number
densities for the various surveys. These numbers could change
depending on the delivered seeing and noise levels for a given
survey. We note that due to PSF degradation effects, 
the effective number density of sources that can used
in the lensing analysis, $n_{\rm eff}$, is lower than the number
density of detected objects. Studies based on simulated and
actual deep ground based images suggest that it is difficult to
exceed an effective density of 30 galaxies arcmin$^{-2}$ in typical
ground based data (this upper limit depends on the seeing and other factors). 
A space based mission is required to reach significantly higher
source densities.

The projects that will start in the immediate future represent a major
step forward, but they will be carried out on mostly general purpose
facilities (with the exception of PanSTARRS). This limits the amount of
time available for large multi-wavelength surveys.  Hence, a
significantly larger, deep survey would require a dedicated large
aperture telescope with a very wide field-of-view. 
One proposal is the expansion of
PanSTARRS to include more telescopes to increase the etendue
(field-of-view times collecting area) of the facility.  A proposal 
for arguably the definitive ground-based survey is the 6-band imaging
survey over 20,000 square degrees proposed for the Large Synoptic Survey
Telescope\footnote{http://www.lsst.org} (LSST), 
an 8.4m telescope with a 10 square degree
field of view  whose survey capacity
is an order of magnitude larger than any Stage III project. 

Alternatively, in an attempt to minimize PSF related systematics,
space based missions are being planned. These have the added benefit that
high quality multi-wavelength data out to NIR wavelenghts can be
obtained.  A European project, the Dark UNiverse
Explorer\footnote{http://www.dune-mission.net} (DUNE) focuses on the
improved image quality and stability that can be obtained from space,
but, in its current form, still relies on extensive ground based
follow-up. Also, the improvement in $n_{\rm eff}$ is relatively small
compared to LSST.  The SuperNova/Acceleration
Probe\footnote{http://snap.lbl.gov} (SNAP) is the most comprehensive
proposal, as it combines stable optics for weak lensing shape
measurements with 9-filter optical/NIR photometry for superb
photometric redshifts. It will survey a smaller area of the
sky than DUNE, but to a greater depth.

\subsection{Prospects for lensing cosmology}

This review has covered weak lensing by galaxies, galaxy
clusters and large-scale structure. We have discussed the measurement
of cosmological parameters via lensing, with a focus on dark matter
and dark energy. The use of lensing tomography for dark energy
measurements can be performed using (two- and three-point) shear correlations, 
galaxy-shear cross-correlations and galaxy clusters. Tests of
the nature of dark matter and of gravity on scales of 10 kpc to
1 Mpc are provided by galaxy and cluster lensing. 
Modified gravity theories that attempt to explain the cosmic
acceleration can also be tested using weak lensing measurements on
larger (cosmological) scales. The success of particular applications
of lensing will no doubt depend on how well systematic errors can be
reduced, corrected from the data or marginalized over in making cosmological
inferences. 

Most weak lensing studies (and therefore this review) have focused on
the measurement of the shear using galaxy images. These studies will
continue to provide unique insights in the dark side of the
universe. In conclusion, we highlight a few other aspects of lensing
that are the subject of ongoing research.  For instance, the lensing
signal can also be inferred by measuring magnification effects which
change the number counts of source galaxies
\cite{Broadhurst95,Scranton07}. These effects can provide a useful
complementary measure of lensing as the systematic errors involved are
quite different from shear measurements.  We mentioned three-point
shear correlations as useful measures of the non-Gaussian distribution
of the lensing mass. Other measures of non-Gaussianity include global
characterization of the topology of lensing maps, topological charge
distributions, peak statistics in convergence maps and so on.

We also discussed the binning of galaxies using photo-z's for lensing
tomography. It is possible to do better and treat the source galaxy
distribution as three-dimensional (with the position in the redshift
direction having much larger uncertainty than in the transverse
direction), which can lead to improved cosmological constraints
\cite{Heavens03}. Another application of tomography is the 
actual reconstruction of the three-dimensional lensing mass
distribution \cite{Taylor01,Hu02}; an attempt has been made using the
COSMOS survey carried out with the HST \cite{Massey07b}.  
A relatively new area of research is the measurement of
higher order derivatives of the lensing potential, which provide
additional information on small scale variations in the mass
distribution \cite{Irwin03,Goldberg05}. 

Finally, we mention two high redshift applications of lensing.  An
area of active study is the effect of lensing by foreground structures
on the CMB \cite{Zaldarriaga98,Hu01}. This is a significant contaminant for
studies of the CMB polarization \cite{Seljak04}, but can also provide
additional information about the lensing mass. Finally, future radio
telescopes may be able to detect galaxies at high redshift through
their 21cm emission. If lensing effects can be measured accurately
with 21cm surveys, they can provide high accuracy power spectra over a
wide range in redshift \cite{Zahn06,Zhang06,Metcalf07}.

{\it Acknowledgement. } We acknowledge helpful
discussions and comments on the manuscript from 
Gary Bernstein, Jacek Guzik, Mike Jarvis, Nick Kaiser, 
Steve Kahn, Eric Linder, Jason Rhodes, Alexandre Refregier, 
Fritz Stabenau, Masahiro Takada, Andy Taylor and Ludo van Waerbeke. 
This work is supported in part by the Department of Energy, the
Research Corporation and NSF grant AST-0607667.

\end{document}